\documentclass[review,12pt,3p]{elsarticle} 
\usepackage{dcolumn}
\usepackage{bm}
\usepackage{graphicx}
\usepackage[dvips]{color}
\usepackage{epsfig}
\usepackage{float}
\usepackage{subfig}
\usepackage{verbatim}
\usepackage{dcolumn}
\usepackage{setspace}
\usepackage{lineno}
\usepackage{rotating} 
\biboptions{compress}

\journal{Phys. Rev. C}

\begin{document}

\title{Neutron-induced $\gamma$-ray production cross sections for the first excited-state transitions in $^{20}$Ne and $^{22}$Ne}

\author[unc,tunl]{S.~MacMullin\corref{cor}} 
\ead{spm@physics.unc.edu}
\author[lanl]{M.~Boswell} 
\author[lanl]{M.~Devlin}
\author[lanl]{S.R.~Elliott}
\author[lanl]{N.~Fotiades}
\author[usd]{V.E.~Guiseppe}
\author[unc,tunl]{R.~Henning}
\author[lanl]{T.~Kawano}
\author[lanl]{B.H.~LaRoque\fnref{fn1}}
\author[lanl]{R.O.~Nelson}
\author[lanl]{J.M.~O'Donnell}

\cortext[cor]{Corresponding author}
\fntext[fn1]{Current address: Department of Physics, University of California, Santa Barbara, CA 93106 USA}
\address[unc]{Department of Physics and Astronomy, University of North Carolina, Chapel Hill, NC 27599 USA}
\address[tunl]{Triangle Universities Nuclear Laboratory, Durham, NC 27708 USA}
\address[lanl]{Los Alamos National Laboratory, Los Alamos, NM 87545 USA}
\address[usd]{Department of Physics, University of South Dakota, Vermillion, SD 57069 USA}


\date{\today}
\begin{abstract}
\textbf{Background:} Neutron-induced reactions are a significant concern for experiments that require extremely low levels of radioactive backgrounds. Measurements of $\gamma$-ray production cross sections over a wide energy range will help to predict and identify neutron backgrounds in these experiments. \textbf{Purpose:} Determine partial $\gamma$-ray production cross sections for neutron-induced reactions in natural neon. \textbf{Methods:} The broad spectrum neutron beam at the Los Alamos Neutron Science Center (LANSCE) was used for the measurement. Gamma rays from neutron-induced reactions were detected using the GErmanium Array for Neutron Induced Excitations (GEANIE). \textbf{Results:} Partial $\gamma$-ray cross sections were measured for the first excited-state transitions in $^{20}$Ne and $^{22}$Ne. The measured cross sections were compared to the TALYS and CoH$_3$ nuclear reaction codes. \textbf{Conclusions:} These are the first experimental data for ($n,n'$) reactions in neon. In addition to providing data to aid in the prediction and identification of neutron backgrounds in low-background experiments, these new measurements will help refine cross-section predictions in a mass region where models are not well constrained. 
\end{abstract}
\clearpage
\maketitle

\section{Introduction}

Several current and next-generation detectors designed to search for WIMP dark matter will make use of large volumes of noble liquids (Ne, Ar, Xe)~\cite{Ber10}. For example, the DEAP/CLEAN experimental program uses either argon or neon~\cite{Bou04,Bou08,Mck07,Him11}. The detectors are designed to measure the scintillation light from putative WIMP-nucleus scattering. Although electrons and $\gamma$ rays, which scatter from atomic electrons, are well-discriminated from nuclear recoils, neutrons, which scatter from nuclei in the detector, will mimic WIMP signals~\cite{Bou06}. These scattering neutrons contribute an irreducible background that has to be quantified using calibration data and Monte Carlo simulations. For the latter, it is crucial that neutron scattering cross sections are well known. 

In nuclear reaction codes used to predict $\gamma$-ray production cross sections, the optical model determines transmission coefficients, which are used in statistical calculations to determine cross sections. The optical model is known to provide an excellent phenomenological description of nucleon-nucleus scattering for medium mass and heavy nuclei ($A > 24$) over a wide energy range ($E < 200$ MeV)~\cite{Hod63}. Global optical models are based on a unique functional form for the energy and mass dependence of the potential depths, with physically constrained parameters that describe nuclear radii and surface diffuseness. The limited range of the global optical-model potentials is largely due to the limited experimental data in the light-to-medium-mass range~\cite{Kon03}. 

We have measured the partial $\gamma$-ray production cross sections for the first excited-state transitions in $^{20}$Ne and $^{22}$Ne for incident neutron energies up to 16 MeV. These are the first experimental data for ($n,n'$) reactions in neon. The inclusion of these cross sections over a wide energy range in Monte Carlo codes will help in predicting neutron backgrounds in dark-matter experiments that use neon as a target. These data will also provide an additional benchmark for global optical-model parameters. This work is a continuation of previous experiments which measured ($n,xn\gamma$) reactions in lead~\cite{Gui09}, copper~\cite{Bos11} and argon~\cite{Mac12}. 

\section{Experiment}

Data were collected at the Los Alamos Neutron Science Center (LANSCE)~\cite{Lis90}. The $\gamma$ rays produced in neutron-induced reactions were measured with the the GErmanium Array for Neutron Induced Excitations (GEANIE)~\cite{Fot04}. GEANIE is located 20.34 m from the Weapons Neutron Research facility (WNR) spallation neutron source on the 60$^{\circ}$-right flight path. A broad-spectrum ($\sim$ 0.2 -- 800 MeV) pulsed neutron beam was produced via spallation on a $^{\rm nat}$W target by an 800-MeV proton linear accelerator beam. The proton beam structure contained 625-$\mu$s long ``macropulses'' repeated at 60 Hz, or every 16.7 ms. One in three  macropules was delivered to another facility, resulting in an average rate of 40 s$^{-1}$. Each macropulse consisted of ``micropulses" spaced every 1.8 $\mu$s, each less than 1 ns long. The pulsed beam allowed incident neutron energies to be determined using the time-of-flight technique. During the experimental runs, $1.2 \times 10^{10}$ micropulses produced about $1.2 \times 10^{12}$ neutrons of energies from 1 to 100 MeV on target to be used in the cross-section analysis. The neutron flux on target was measured with an in-beam fission ionization chamber with $^{235}$U and $^{238}$U foils~\cite{Wen93} located about two meters upstream from the center of the GEANIE array.

GEANIE comprises 20 high-purity germanium detectors with bismuth germinate (BGO) escape suppression shields. Detectors are either a planar or coaxial geometry and are typically operated with maximum $\gamma$-ray energy ranges of 1 MeV and 4 MeV, respectively. Since almost all of the excited states in neon produce $\gamma$ rays with energies greater than 1 MeV, the planar detectors were removed from the data processing chain to reduce dead time. The data from five out of nine of the the coaxial detectors with good energy resolution and timing information were used for the cross-section analysis. 

The neon gas target cell was a 3.81-cm diameter and 6.35-cm length thin-walled aluminum cylinder with 0.127-mm thick Kapton windows at either end. The gas cell was placed at the center of the GEANIE array, with the neutron beam passing through the Kapton foils. The $^{\rm nat}$Ne gas pressure was maintained at 3.96 atm with less than 1\% variation over the course of the experiment.

\section{Experimental results and discussion}

\begin{figure}[htp]
\centering
\includegraphics[width=0.95\textwidth, angle=0]{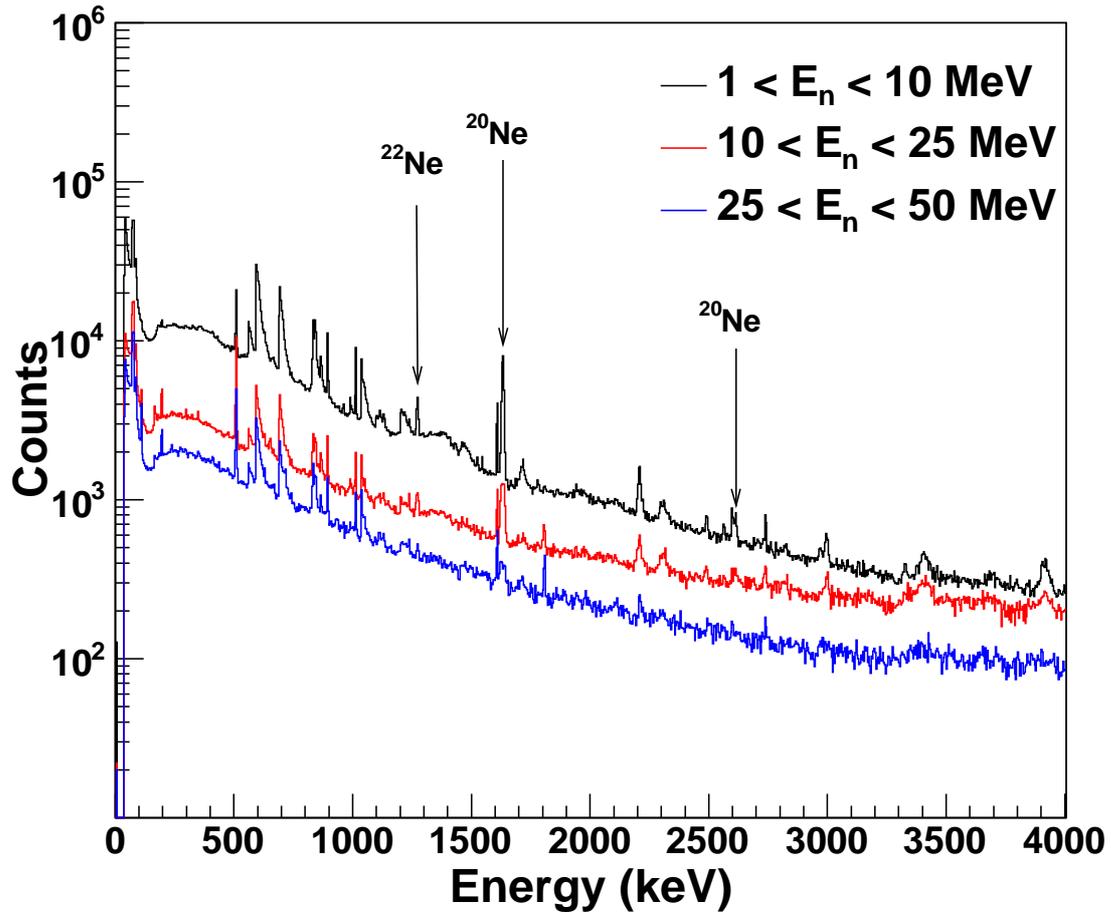}
\caption{Neon-sample $\gamma$-ray spectra selected for different neutron energy windows. The spectrum shown in black (top) corresponds to $1 < E_n < 10$ MeV. The spectrum shown in red (middle) corresponds to $10 < E_n < 25$ MeV. The spectrum shown in blue (bottom) corresponds to $25 < E_n < 50$ MeV. The observed excited-state transitions in $^{20}$Ne and $^{22}$Ne are labeled.}
\label{fig:neonspectra}
\end{figure}

Partial $\gamma$-ray cross sections were determined using the same method described in Ref.~\cite{Mac12}. Neon-sample $\gamma$-ray spectra selected for specific neutron energy windows are shown in Fig.~\ref{fig:neonspectra}. All $\gamma$-ray lines present in the data have been identified. Partial $\gamma$-ray cross sections were determined for the $E_\gamma = 1633.7$-keV $2^+ \rightarrow 0$ transition in $^{20}$Ne and the $E_\gamma = 1274.5$-keV $2^+ \rightarrow 0$ transition in $^{22}$Ne. The $E_\gamma = 2613.8$-keV $4^+ \rightarrow 2$ transition in $^{20}$Ne was visible in the data, but only over a broad energy bin. Most of the other $\gamma$-ray lines were attributed to backgrounds from the sample cell ($^{27}$Al) or neutron inelastic scattering in germanium or bismuth (from the BGO shields). 

\begin{centering}
\begin{figure}
  \centering
  \subfloat{\label{fig:ne20}\includegraphics[width=0.85\textwidth]{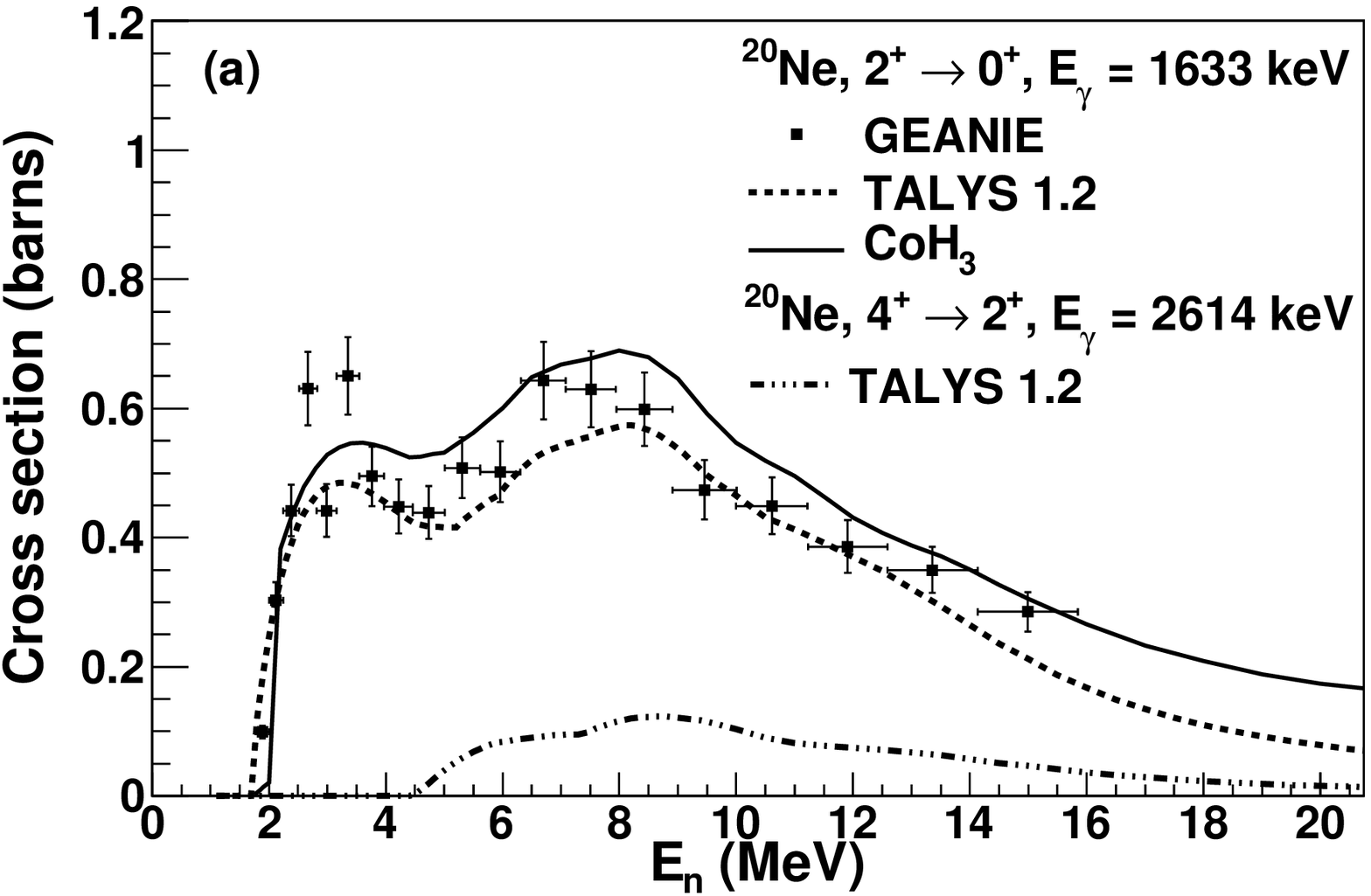}}\\
  \subfloat{\label{fig:ne22}\includegraphics[width=0.85\textwidth]{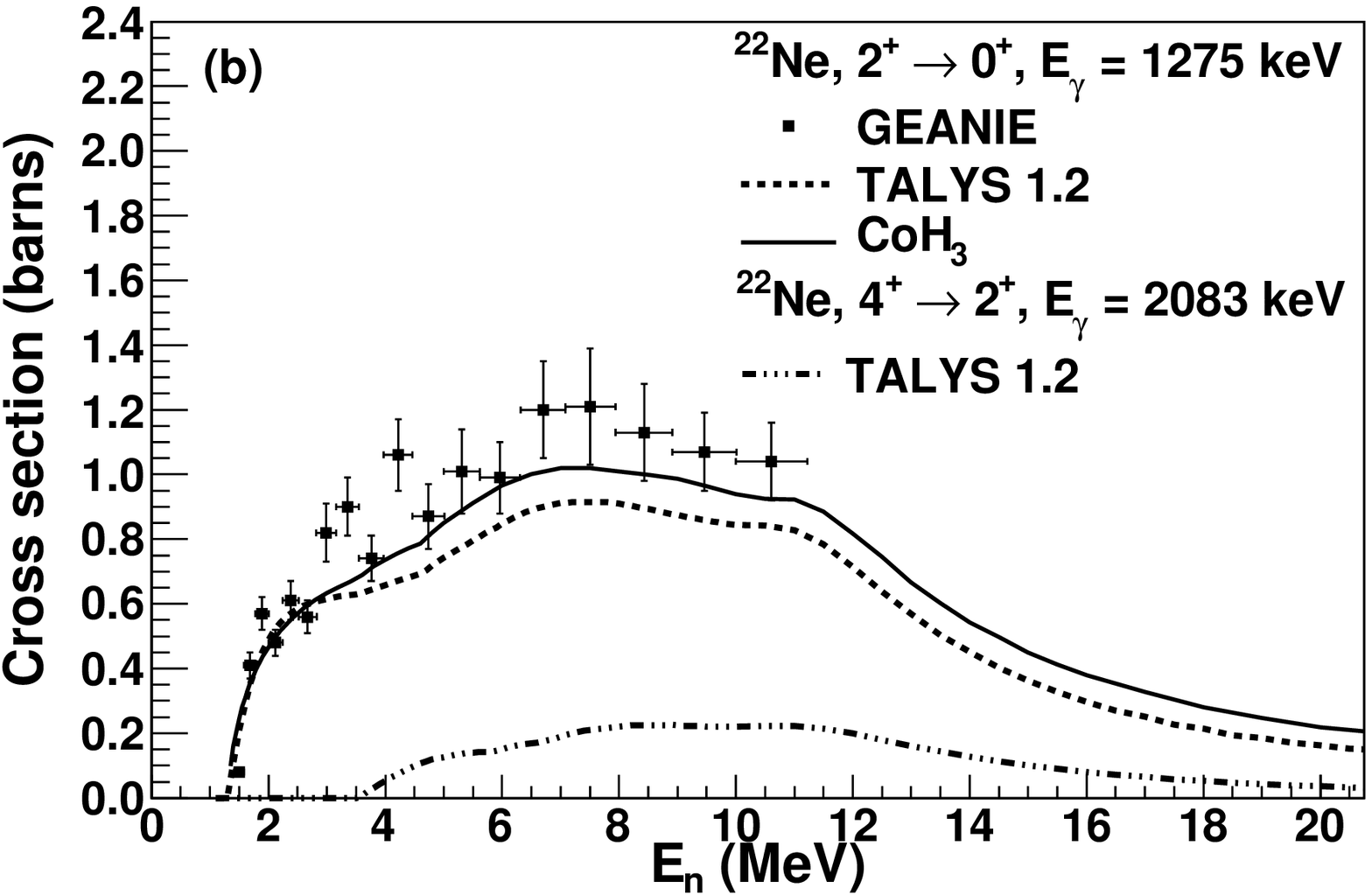}}                 
  \caption{(a) Partial $\gamma$-ray cross section for $^{20}$Ne. The data from the $2^{+} \rightarrow 0^{+}$ first excited-state transition is compared with the TALYS and CoH$_3$ calculations. A TALYS calculation for the $4^{+} \rightarrow 2^{+}$ second excited-state transition is also shown. (b) Partial $\gamma$-ray cross section for $^{22}$Ne. The data from the $2^{+} \rightarrow 0^{+}$ first excited-state transition is compared with the TALYS and CoH$_3$ calculations. A TALYS calculation for the $4^{+} \rightarrow 2^{+}$ second excited-state transition is also shown.}
  \label{fig:neon}
\end{figure}
\end{centering}

We assumed that the observed $\gamma$ ray at 1633 keV was due only to the $^{20}$Ne($n,n'\gamma$)$^{20}$Ne reaction. Considering that neon is 90.48\% $^{20}$Ne, 9.25\% $^{22}$Ne and 0.27\% $^{21}$Ne, the only other reactions that could produce $^{20}$Ne are $^{21}$Ne($n,2n\gamma$)$^{20}$Ne or $^{22}$Ne($n,3n\gamma$)$^{20}$Ne. The low isotopic abundance of $^{21}$Ne makes the $^{21}$Ne($n,2n\gamma$)$^{20}$Ne reaction unlikely to be seen. The $^{22}$Ne($n,3n\gamma$)$^{20}$Ne does not contribute to the measured cross section because it has a threshold of more than 20 MeV. Similarly, the 1275-keV transition observed from $^{22}$Ne was assumed to be from the $^{22}$Ne($n,n'\gamma$)$^{22}$Ne reaction. Cross sections were measured from threshold to where they fall below our detection sensitivity. The measured cross sections are shown in Fig.~\ref{fig:neon}.

The TALYS~\cite{Kon05} and CoH$_3$~\cite{Kaw03,Kaw10} nuclear reaction codes were used to predict the $\gamma$-ray production cross sections for the transitions studied in the present work. The TALYS calculation used a direct reaction model using the global optical-model parameterization of Koning and Delaroche~\cite{Kon03}, a pre-equilibrium model and a Hauser-Feshbach statistical calculation. Because neon is a deformed nucleus ($\beta_2=0.73$ for $^{20}$Ne and $\beta_2=0.63$ for $^{22}$Ne), the CoH$_3$ code used a Hauser-Feshbach calculation, including transmission coefficients obtained from a coupled-channels calculation~\cite{Kaw09}. The discrepancy between the data and CoH$_3$ calculation is most likely due to the use of the spherical Koning-Delaroche potential in the coupled-channels calculation. The peaks in the $^{20}$Ne cross section around 3 MeV are due to resonances. Neither the TALYS nor the CoH$_3$ calculations reproduce the structure, but rather give the average behavior of the cross section.

The fact that only the first excited-state transitions in $^{20}$Ne and $^{22}$Ne were observed in the data has potential implications for neon-based dark matter experiments. Although neutron elastic scattering can mimic a WIMP signal, experiments may be able to discriminate against inelastic scattering if one or more coincident $\gamma$ rays are produced. While the cross sections for the first excited-state transitions in $^{20}$Ne and $^{22}$Ne are relatively large, TALYS calculations, shown in Fig.~\ref{fig:neon}, indicate that the $\gamma$-ray production cross sections from higher excited-state transitions in both $^{20}$Ne and $^{22}$Ne are at least a factor of five smaller for neutron energies less than about 10 MeV. 

\begin{centering}
\begin{figure}
  \centering             
  \includegraphics[width=0.95\textwidth]{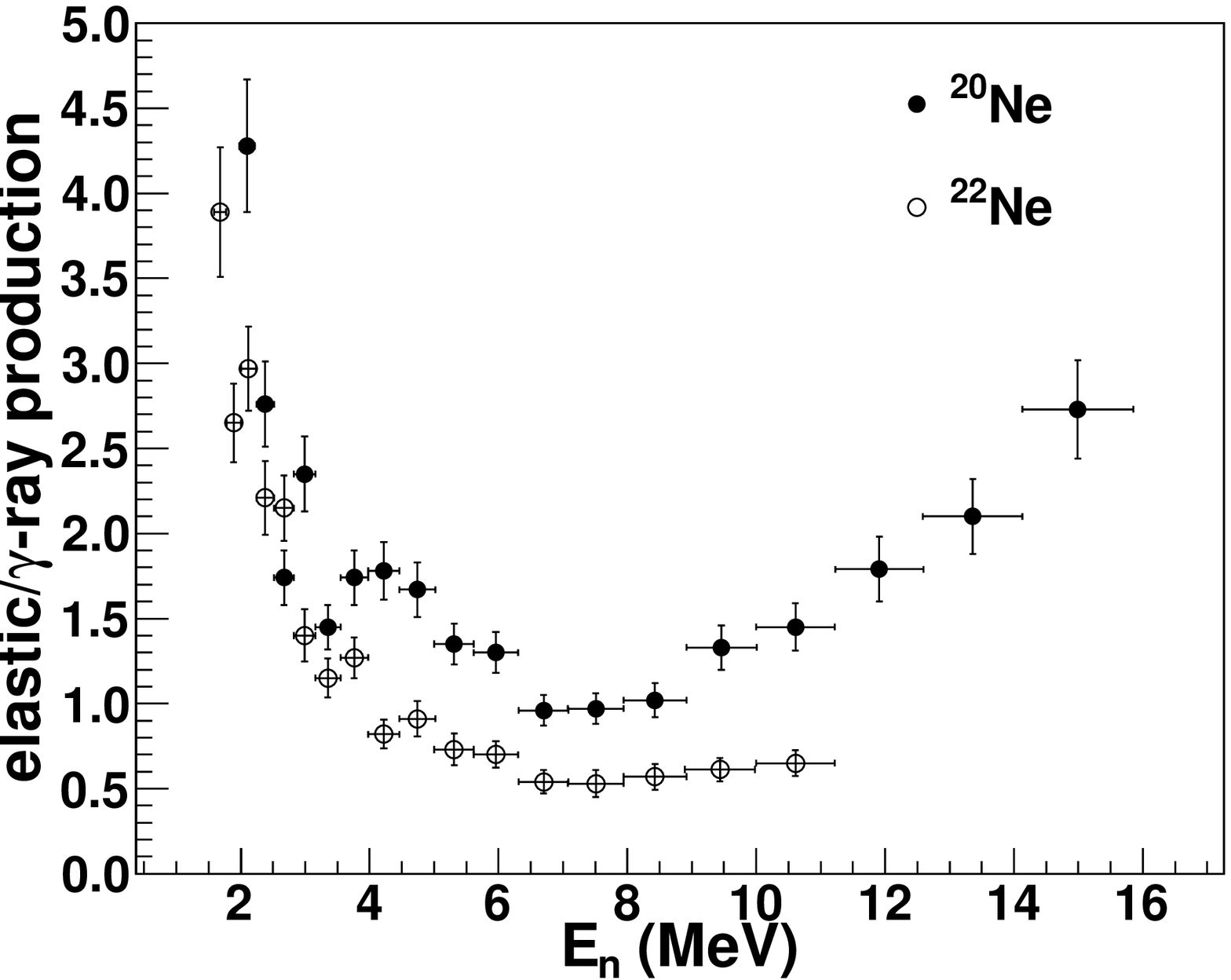}
  \caption{The ratio of the elastic scattering cross section to the $\gamma$-ray production (inelastic) cross section. The solid circles correspond to $^{20}$Ne and the open circles correspond to $^{22}$Ne.}
  \label{fig:elinelcompare}
\end{figure}
\end{centering}

In addition to using neutron inelastic scattering as an active veto, if the rate of inelastic scattering in the detector's sensitive volume can be measured, the background from neutron elastic scattering may be estimated if both the elastic and $\gamma$-ray production cross sections are known. In most underground experiments, the dominant source of neutrons are those produced from naturally occurring isotopes in the $^{238}$U and $^{232}$Th decay chains. Alpha particles produced in these decays undergo ($\alpha,n$) reactions and produce neutrons in the 1~to~10~MeV range. The ratio of the elastic scattering to $\gamma$-ray production cross sections for $^{20}$Ne and $^{22}$Ne are shown in Fig.~\ref{fig:elinelcompare}. The elastic scattering cross section was calculated using TALYS, which used the Koning-Delaroche global optical model including a compound nucleus cross section using a Hauser-Feshbach statistical calculation with a Moldauer width fluctuation correction factor~\cite{Mol76,Mol80}. The $\gamma$-ray production cross section corresponds to the values measured in the current experiment. Since there is currently no experimental data for elastic scattering in neon to use for comparison, the error bars only represent the current experiment. Although the ratio of the cross sections becomes large as the neutron energy approaches threshold, only about 15\% of the total neutrons produced from $^{238}$U and $^{232}$Th-induced ($\alpha,n$) reactions have energies below 2 MeV~\cite{Mei09}.
 
\section{Summary}

We have measured neutron-induced $\gamma$-ray production cross sections for the first excited-state transitions in $^{20}$Ne and $^{22}$Ne from threshold to as high as 16 MeV, where they fall below our detection sensitivity. The measured cross sections will aid in the identification and discrimination of neutrons in underground experiments which use neon as a detector. Since these are the first experimental data for ($n,n'$) reactions in neon, they will enrich the nuclear databases and provide a useful benchmark in a mass region where the optical model is not well constrained. 
 
\section{Acknowledgements}
This work was supported in part by Laboratory Directed Research and Development at Los Alamos National Laboratory, National Science Foundation Grant 0758120 and US Department of Energy grant number 2013LANLE9BW. This work benefited from the use of the Los Alamos Neutron Science Center, funded by the US Department of Energy under Contract DE-AC52-06NA25396. Henning and MacMullin are supported by DOE ONP grant awards DE-FG02-97ER4104 and DE-FG02-266 97ER41033. Guiseppe is supported by DOE ONP grant number is DE-SCOO05054. 
\clearpage


\clearpage
\appendix
\section{Partial $\gamma$-ray Cross Sections}

\begin{table}[htp]
	\centering
	\caption{$^{20}$Ne($n,n'\gamma$)$^{20}$Ne  $2^{+} \rightarrow 0^{+}$  $E_\gamma = 1633$ keV }
	\begin{tabular}{l c c c}
\hline
\hline
$E_n$ (MeV)	&	$\sigma_{\rm data}$ (barn) & $\sigma_{\rm TALYS}$ (barn) & $\sigma_{\rm CoH_3}$ (barn)\\  
\hline
1.9 $\pm$ 0.1	& 0.10	$\pm$ 0.01   &   0.19    & 0.13 \\
2.1	$\pm$ 0.1	& 0.30	$\pm$ 0.03    &   0.31    & 0.30 \\
2.4	$\pm$ 0.1	& 0.44	$\pm$ 0.04    &   0.39    & 0.44 \\
2.7	$\pm$ 0.2	& 0.63	$\pm$ 0.06    &   0.45    & 0.49\\
3.0	$\pm$ 0.2	& 0.44	$\pm$ 0.04    &   0.48    & 0.53 \\
3.4	$\pm$ 0.2	& 0.65	$\pm$ 0.06    &   0.49    & 0.55 \\
3.8	$\pm$ 0.2	& 0.50	$\pm$ 0.05    &   0.47    & 0.54 \\
4.2	$\pm$ 0.2	& 0.45	$\pm$ 0.04    &   0.44    & 0.53 \\
4.7	$\pm$ 0.3	& 0.44	$\pm$ 0.04    &   0.42    & 0.53 \\
5.3	$\pm$ 0.3	& 0.51	$\pm$ 0.05    &   0.43    & 0.56 \\
6.0	$\pm$ 0.3	& 0.50	$\pm$ 0.05    &   0.48    & 0.60 \\
6.7	$\pm$ 0.4	& 0.64	$\pm$ 0.06    &   0.53    & 0.66 \\
7.5	$\pm$ 0.4	& 0.63	$\pm$ 0.06    &   0.56    & 0.68 \\
8.4	$\pm$ 0.5	& 0.60	$\pm$ 0.06    &   0.57    & 0.68 \\
9.5	$\pm$ 0.5	& 0.47	$\pm$ 0.05    &   0.50    & 0.59 \\
10.6 $\pm$ 0.6 & 0.45	$\pm$ 0.04    &   0.43    & 0.52 \\
11.9 $\pm$ 0.7 & 0.39	$\pm$ 0.04    &   0.38    & 0.43 \\
13.4 $\pm$ 0.7 & 0.35	$\pm$ 0.04    &   0.31    & 0.37 \\
15.0 $\pm$ 0.8 & 0.29   $\pm$ 0.03    &   0.22    & 0.31 \\
\hline
\hline
\end{tabular}
\label{tab:geanie_neon_1633}
\end{table}

\begin{table}[htp]
	\centering
	\caption{$^{22}$Ne($n,n'\gamma$)$^{22}$Ne  $2^{+} \rightarrow 0^{+}$  $E_\gamma = 1275$ keV }
	\begin{tabular}{l c c c}
\hline
\hline
$E_n$ (MeV)	&	$\sigma_{\rm data}$ (barn) & $\sigma_{\rm TALYS}$ (barn) & $\sigma_{\rm CoH_3}$ (barn)\\   
\hline
1.5	$\pm$ 0.1 &	0.08 $\pm$ 0.01 & 0.00 & 0.25 \\
1.7	$\pm$ 0.1 &	0.41 $\pm$ 0.04 & 0.34 & 0.36 \\
1.9 $\pm$ 0.1 &	0.57 $\pm$ 0.05 & 0.46 & 0.44 \\
2.1	$\pm$ 0.1 &	0.48 $\pm$ 0.04 & 0.53 & 0.49 \\
2.4	$\pm$ 0.1 &	0.61 $\pm$ 0.06 & 0.58 & 0.55 \\
2.7 $\pm$ 0.2 &	0.56 $\pm$ 0.05 & 0.61 & 0.60 \\
3.0 $\pm$ 0.2 &	0.82 $\pm$ 0.09 & 0.63 & 0.63 \\
3.4	$\pm$ 0.2 &	0.90 $\pm$ 0.09 & 0.63 & 0.67 \\
3.8	$\pm$ 0.2 &	0.74 $\pm$ 0.07 & 0.65 & 0.72 \\
4.2	$\pm$ 0.2 &	1.06 $\pm$ 0.11 & 0.65 & 0.78 \\
4.7	$\pm$ 0.3 &	0.87 $\pm$ 0.10 & 0.71 & 0.80 \\
5.3	$\pm$ 0.3 &	1.01 $\pm$ 0.13 & 0.78 & 0.90 \\
6.0	$\pm$ 0.3 &	0.99 $\pm$ 0.11 & 0.85 & 0.97 \\
6.7	$\pm$ 0.4 &	1.20 $\pm$ 0.15 & 0.91 & 1.02 \\
7.5	$\pm$ 0.4 &	1.21 $\pm$ 0.18 & 0.92 & 1.02 \\
8.4	$\pm$ 0.5 &	1.13 $\pm$ 0.15 & 0.90 & 1.00 \\
9.5	$\pm$ 0.5 &	1.07 $\pm$ 0.12 & 0.86 & 0.96 \\
10.6 $\pm$ 0.6 & 1.04 $\pm$ 0.12 & 0.84 & 0.92 \\
\hline
\hline
\end{tabular}
\label{tab:geanie_neon_1275}
\end{table}

\begin{table}[htp]
	\centering
	\caption{The ratio of the elastic scattering cross section to the measured $\gamma$-ray production (inelastic) cross section for $^{20}$Ne and $^{22}$Ne.}
	\begin{tabular}{l c c}
\hline
\hline
$E_n$ (MeV)	&	$^{20}$Ne & $^{22}$Ne \\   
\hline
1.5	$\pm$ 0.1 &	 & 21.2 $\pm$ 2.6 \\
1.7	$\pm$ 0.1 &	 &  3.9 $\pm$ 0.4 \\
1.9 $\pm$ 0.1	& 13.9	$\pm$ 1.4   &   2.7 $\pm$ 0.2 \\
2.1	$\pm$ 0.1	& 4.3	$\pm$ 0.4    &   3.0 $\pm$ 0.2 \\
2.4	$\pm$ 0.1	& 2.8	$\pm$ 0.3    & 2.2 $\pm$ 0.2 \\
2.7	$\pm$ 0.2	& 1.7	$\pm$ 0.2    &  2.2 $\pm$ 0.2 \\
3.0	$\pm$ 0.2	& 2.4	$\pm$ 0.2    &  1.4 $\pm$ 0.2 \\
3.4	$\pm$ 0.2	& 1.5	$\pm$ 0.1    &   1.2 $\pm$ 0.1 \\
3.8	$\pm$ 0.2	& 1.7	$\pm$ 0.2    &   1.3 $\pm$ 0.1 \\
4.2	$\pm$ 0.2	& 1.8	$\pm$ 0.2    &   0.82 $\pm$ 0.08 \\
4.7	$\pm$ 0.3	& 1.7	$\pm$ 0.2    &   0.9 $\pm$ 0.1 \\
5.3	$\pm$ 0.3	& 1.4	$\pm$ 0.1    &   0.73 $\pm$ 0.09 \\
6.0	$\pm$ 0.3	& 1.3	$\pm$ 0.1    &   0.70 $\pm$ 0.08 \\
6.7	$\pm$ 0.4	& 0.96	$\pm$ 0.09    &  0.54 $\pm$ 0.07 \\
7.5	$\pm$ 0.4	& 0.97	$\pm$ 0.09    &  0.53 $\pm$ 0.08  \\
8.4	$\pm$ 0.5	& 1.0	$\pm$ 0.1    &  0.57 $\pm$ 0.08  \\
9.5	$\pm$ 0.5	& 1.3	$\pm$ 0.1    &   0.61 $\pm$ 0.07 \\
10.6 $\pm$ 0.6 & 1.5	$\pm$ 0.1    &  0.65 $\pm$ 0.07 \\
11.9 $\pm$ 0.7 & 1.8 $\pm$ 0.2    &    \\
13.4 $\pm$ 0.7 & 2.1	$\pm$ 0.2    &    \\
15.0 $\pm$ 0.8 & 2.7   $\pm$ 0.3    &    \\
\hline
\hline
\end{tabular}
\label{tab:ratio}
\end{table}

\end{document}